\documentstyle[11pt,aaspp4,epsf,flushrt]{article}
\begin{document}

\font\BF=cmmib10 scaled 1200
\def\k{\hbox{\BF k}}
\def\khat{\hat{\k}}
\def\om{\Omega_m}
\def\ov{\Omega_\Lambda}
\def\go{\gtrsim}
\def\lo{\lesssim}
\def\deltaD{\delta_{\rm D}}
\def\eq{{\rm eq}}

%%%%%%%%%%%%%%%%%%%%%%%%%%%%%%%%%%%%%%%%%%%%%%%%%%%%%%%%%%%%%%%%%%%%%%%%%
\title{Halo Profiles and the Nonlinear Two- and Three-Point Correlation 
Functions of Cosmological Mass Density}

\author{Chung-Pei Ma}
\affil{Department of Physics and Astronomy,
University of Pennsylvania, Philadelphia, PA 19104; 
cpma@strad.physics.upenn.edu}
\and
\author{J. N. Fry}
\affil{Department of Physics, University of Florida, 
Gainesville, FL 32611-8440; fry@phys.ufl.edu}

%%%%%%%%%%%%%%%%%%%%%%%%%%%%%%%%%%%%%%%%%%%%%%%%%%%%%%%%%%%%%%%%%%%%%%%%%
\begin{abstract}

We investigate the nonlinear two- and three-point correlation
functions of the cosmological density field in Fourier space and test
the popular hierarchical clustering model, that the three-point
amplitude $Q$ is independent of scale.  In high-resolution $N$-body
simulations of both $n=-2$ scale-free and cold dark matter models, we
find that $Q$ at late times is not constant but increases with
wavenumber far into the nonlinear regime.  Self-similar scaling also
does not hold as rigorously for the three-point function as for the
two-point function in the $n=-2$ simulation, perhaps a manifestation
of the finite simulation volume.
We suggest that a better understanding of the two- and three-point
correlation functions in the nonlinear regime lies in the link to the
density profiles of dark matter halos.  We demonstrate and quantify
how the slopes of the correlation functions are affected by the slope
of the halo profiles using simple halo shapes and analytic clustering
models.

\end{abstract}


%%%%%%%%%%%%%%%%%%%%%%%%%%%%%%%%%%%%%%%%%%%%%%%%%%%%%%%%%%%%%%%%%%%%%%%%%
\section{Introduction}

In this {\it Letter} we examine the power spectrum and the bispectrum
for numerical simulations of the $n = -2$ scale free model and cold
dark matter (CDM) model.  The power spectrum $P(k)$ is the most
fundamental statistic for a random field and the only statistic for a
Gaussian field.  The cosmological distribution of matter, however, is
not Gaussian, and higher order statistics are also important.  The
lowest-order non-Gaussian information is encoded in the bispectrum
$B(k_1,k_2,k_3)$.  In analogy to $P(k)$, which is the Fourier
transform of the two-point correlation function $\xi\,$, the
bispectrum is the Fourier transform of the three-point correlation
function $\zeta\,$.  The power spectrum $P(k)$ of matter fluctuations
is related to the mass density field $\delta=\rho/\bar{\rho}-1$ in
$k$-space by $\langle \delta(\k_1) \delta(\k_2) \rangle = P(k_1) \,
\deltaD (\k_1 + \k_2)\,$, where $ \deltaD $ is the Dirac
delta-function.  The bispectrum $B(k_1,k_2,k_3)$, similarly, is
related to $\delta$ by $ \langle \delta(\k_1) \delta(\k_2)
\delta(\k_3) \rangle = B(k_1,k_2,k_3)\, \deltaD (\k_1 +\k_2 +\k_3)\,$.

On sufficiently large length scales, density fluctuations are small
enough that perturbation theory for gravitational instability
suffices, and the predictions for the correlation functions can be
computed in detail.  To linear order, the fluctuation amplitude
$\delta(\k,t)$ grows by an overall scale factor,
$\delta=D(t)\delta_0(\k)$, where $\delta_0(\k)$ is the amplitude at
some early time $t_0$ and $D(t)$ is a growing function of time.  In a
nonlinear theory, interactions generate a sum of terms to all orders
in the initial amplitude, $ \delta = \delta^{(1)} + \delta^{(2)} +
\delta^{(3)} + \cdots\,$, where $ \delta^{(n)}\sim \delta_0^{\,n} $.
Even for a Gaussian initial distribution, these nonlinear terms induce
nonvanishing higher order correlations.  In perturbation theory, the
lowest order two-point statistic is the linear power spectrum $P_l(k)
\propto [\delta^{(1)}]^2$.  For the three-point statistic, the first
nonvanishing contribution (i.e. tree-level) to the bispectrum is
related to $P_l$ by
\begin{equation}
	B^{(0)}(k_1,k_2,k_3)= F(\k_1 ,\k_2) \,P_l(k_1) P_l(k_2) + 
	F(\k_2 ,\k_3) \, P_l(k_2) P_l(k_3) + 
	F(\k_3, \k_1) \, P_l(k_3) P_l(k_1)\,,
\label{btree}
\end{equation}
where $ F(\k_i,\k_j) = {10 \over 7} + (k_i/k_j + k_j/k_i)\,
(\khat_i\cdot\khat_j) + {4 \over 7} (\khat_i \cdot \khat_j)^2 \,$ (Fry
1984).  For $ \Omega < 1 $ the factors $ {10\over 7} $ and $ {4 \over
7} $ become $ 1 \pm \kappa $, where for $ \Omega $ near 1$, \kappa
\approx {3 \over 7}\,\Omega ^{-2/63} $ in an open universe and $
\kappa \approx {3 \over 7}\,\Omega ^{-1/143} $ in a flat universe with
cosmological constant (Bouchet et al.~1995).

Equation~(\ref{btree}) indicates that the main dependence on the power
spectrum can be removed if one defines the hierarchical three-point
amplitude
\begin{equation}
	Q(k_1,k_2,k_3) \equiv { B(k_1,k_2,k_3) \over P(k_1) P(k_2) +
	P(k_2) P(k_3) + P(k_3) P(k_1)} \,.
\label{Qdef}
\end{equation} 
The three-point amplitude $Q$ has the convenient feature that for the
lowest nonvanishing result in perturbation theory (eq.~[\ref{btree}]),
$Q^{(0)}$ is independent of time and the overall amplitude of $P_l$.
Moreover, for scale-free models with a power-law $P_l(k)$, $Q^{(0)}$
is independent of overall scale as well.  In a pure hierarchical
model, $Q$ is exactly constant.

The bispectrum $B$ and the three-point amplitude $Q$ depend on any
three parameters that define a triangle in $k$-space.  For simplicity,
we will use equilateral triangles to explore scale dependence.  We
have examined several other configurations and found behavior similar
to the equilateral case.  For a closed equilateral triangle, one has $
k_1=k_2=k_3 = k$ and $\khat_i\cdot\khat_j = -{1 \over 2} $, and the
bispectrum $B_\eq $ depends only on the single wavenumber $k$.  To 
lowest order, the equilateral bispectrum has a particularly simple
form, $B^{(0)}_\eq (k) = {12\over 7}\, P_l^2(k)\,$, and it follows
from equation~(\ref{Qdef}) that $Q^{(0)}_\eq (k)={4\over 7}\,$, 
independent of the power spectrum.  Since $B$ has dimensions of $
{\rm Mpc}^6 $, it is convenient to define a new dimensionless
variable, which we name $\Delta_3$, to characterize the three-point
statistic of mass density fluctuations.  Following the familiar
density variance $\Delta(k)=4\pi k^3\, P(k)$ for the two-point
statistic, we define $\Delta_3$ for equilateral triangles as
\begin{equation}
	\Delta_3(k) \equiv 4\pi k^3\, \sqrt{B_\eq (k)} \, ;
\label{del3}
\end{equation}
and it follows that $\Delta_3(k)/\Delta(k)=\sqrt{3\,Q_\eq(k)}$.
%\begin{equation}
%{\Delta_3(k) \over \Delta(k)} = \sqrt{3\,Q_\eq(k)} \,.
%\end{equation}
Since $Q_\eq $ is simply a constant to the lowest nonvanishing order
in perturbation theory, this new statistic also has the convenient
property that the leading order $\Delta^{(0)}_3(k)$ and
$\Delta^{(0)}(k)$ have identical shapes, i.e.,
$\Delta^{(0)}_3(k)/\Delta^{(0)}(k)=(12/7)^{1/2}$.

%%%%%%%%%%%%%%%%%%%%%%%%%%%%%%%%%%%%%%%%%%%%%%%%%%%%%%%%%%%%%%%%%%%%%%%%%
\section{High-$k$ Behavior}

The leading-order perturbative expressions given above are valid for $
\delta \ll 1$, or at low $k$.  Including the next-to-leading order terms
for the bispectrum extends the range of validity to the quasilinear regime 
$\delta\sim 1$ (Scoccimarro et al. 1998).  Numerical simulations,
however, are required in order to determine the fully evolved
bispectrum of mass density fluctuations into the deeply nonlinear regime
at high $k$.

We examine results from two cosmological simulations, an $n=-2$
scale-free model and a cold dark matter (CDM) model.  The $n=-2$
simulation has $256^3$ particles and a Plummer force softening length
of $L/5120$, where $L$ is the box length.  This is the highest
resolution simulation of an $n=-2$ model to our knowledge.  It is also
the main simulation used in the study of self-similar scaling of the
power spectrum by Jain \& Bertschinger (1998).  The CDM model is
critically flat with matter density $\om=0.3$ and cosmological
constant $\ov=0.7$.  This run has $128^3$ particles and is performed
in a $ (100\,{\rm Mpc})^3$ comoving box with a comoving force
softening length of $ 50\,{\rm kpc} $ for Hubble parameter $ h=0.75 $.
The baryon fraction is set to zero for simplicity.  The primordial
power spectrum has a spectral index of $n=1$, and the density
fluctuations are drawn from a random Gaussian distribution.  The
gravitational forces are computed with a particle-particle
particle-mesh (P$^3$M) code (Ferrell \& Bertschinger 1994).  We
compute the density field $\delta$ on a grid from particle positions
using the second-order triangular-shaped cloud (TSC) interpolation
scheme.  A fast Fourier transform is then used to obtain $\delta$ in
$k$-space.  The $k$-space TSC window function is deconvolved to
correct for smearing in real space due to the interpolation, and shot
noise terms are subtracted to correct for discreteness effects.

Figure~1 illustrates the behavior of the nonlinear power spectrum
$\Delta(k)$ and the three-point statistics $\Delta_3(k)$ and $Q(k)$
for equilateral triangular configurations for the $n=-2$ scale-free
simulation.  Three time outputs are shown, and the results are plotted
against scaled $ k/k_{nl}$, where $k_{nl}$ is defined by
$\int_0^{k_{nl}} d^3k\, P_l(a,k)=1$, and the expansion factor $a$ is 1
initially.  Unlike the two-point $\Delta$, for which self-similar
scaling works well (Jain \& Bertschinger 1998), the three-point curves
in the plot do not overlap; self-similarity therefore does not 
hold as rigorously for the three-point function in the simulations.
The slope of $Q$ also has a weak dependence on time.  
For the last output ($a=26.91$,
$k_{nl}=7.25$), $Q$ increases monotonically with $k$ down to the
resolution limit with an approximate logarithmic slope of $0.225$.
The next-to-leading order perturbative results for $Q$
(Scoccimarro et al. 1998) roughly agree with the most weakly evolved,  
($a=13.45$, $k_{nl}=29$) 
output up to $k\approx k_{nl}$, but not at higher $k$ nor
in the later outputs.  The upturn of $Q_{\rm eq}$ at high $k$
(dotted portion of curves in Fig.~1) is related to the downturn in
$\Delta$ at the same $k$, which is likely due to limited numerical
resolution.

Many earlier papers did not report this behavior (e.g., Efstathiou et
al.~1988; Fry, Melott, \& Shandarin 1993; Scoccimarro et al.~1998).
Some of the computations were not carried out to sufficiently high
wavenumbers to provide a clear test of the hierarchical model.
Others (e.g. Scoccimarro et al. 1998) {\it assumed} self-similarity
and averaged results over different time outputs, treating each output
as if it were from an independent realization.  We have checked for
possible numerical artifacts by re-examining the $n=-2$ results of Fry
et al. (1993) from $128^3$ particle-mesh (PM) simulations.  At
evolution stages comparable to those in Figure~1 (i.e. $k_{nl}=32$,
16, and 8), we find that $Q_{\rm eq}$ in four different random
realizations agrees well with Figure~1 up to the resolution limit of
the PM runs.

Departures from the hierarchical model have been suggested in several
papers that have examined the real-space three-point function (Suto \&
Matsubara 1994) and the skewness $S_3$ (e.g., Lahav et al.~1993;
Colombi et al.~1996), a volume integral average of the full
three-point function.  The simulations used in these papers had $64^3$
particles and a smaller dynamic range, so the results were limited to
lower $k$.  The skewness is also a less powerful probe of hierarchical
clustering than the full three-point examined here because the volume
integral might average out non-hierarchical signals.

%%%%%%%%%%%%%%%%%%%%%%%%%%%%%%%%%%%%%%%%%%%%%%%%%%%%%%%%%%%%%%%%%%%%%%%%%
\section{Dependence on Halo Density Profiles}

In this section we examine analytical models that can be used to
elucidate the numerical results of Figure~1.  We suggest that the key
to a better understanding of $Q$ in the high-$k$ regime lies in the
connection between $Q$ and the density profiles of dark matter halos.
This is because the contributions to $Q$ at high $k$ are dominated by
close particle triplets, many of which are located within individual
halos of size $\sim 2\pi/k$.  It is therefore reasonable to expect the
two- and three-point correlation functions on small length scales to
reflect the density profiles of dark matter halos.  Sheth and Jain
(1997) have examined how the amplitude of the two-point function is
related to simple halo profiles.  Our emphasis here is on the link
between the shape of the three-point function and halo profiles.

One simple way to demonstrate the relation between halos and the
correlation functions is to consider the ``power law cluster'' model.
In this model, matter is distributed in randomly placed halos with a
power-law mass density profile $\rho(r) \propto r^{-\epsilon}$ out to
some radius $R$ (Peebles 1974; 1980).  For $1.5< \epsilon < 3$ and
$r<R$, one can derive analytically that the two-point function $\xi$,
the density variance $\Delta$, the three-point function $\zeta$, and
$Q$ scale as
\begin{equation}
	\xi(r)\propto r^{3-2\epsilon}\,, \quad 
	\Delta(k) \propto k^{2\epsilon-3}\,, \quad 
	\zeta(r)\propto r^{3-3\epsilon}\,, \quad
	Q \propto {\zeta\over \xi^2} \propto r^{\epsilon-3} 
	\propto k^{3-\epsilon}\,. \label{epsilon}
\end{equation}
This model is clearly simplistic because in realistic cosmological
models, the centers of dark matter halos are not randomly distributed,
the halo density profile is not a pure power law, and not all mass is
located within the halos.  Equation~(\ref{epsilon}) nonetheless
provides valuable insight into the link between the shape of dark
matter halos and the high-$k$ behavior of the two- and three-point
clustering statistics.

To test the applicability of equation~(\ref{epsilon}) in a more
realistic setting, we have performed a series of experiments with dark
matter halos identified in cosmological $N$-body simulations.  In
these experiments, we keep the centers and masses of the halos
unchanged but redistribute the subset of particles which lies within
the virial radius (the radius within which $\delta > 200$) of
each halo according to $\rho\propto r^{-\epsilon}$ for a chosen
$\epsilon$.  We then recompute $\Delta$ and $ Q_\eq $ from the
particle positions.  This recipe allows us to explore simple density
profiles while preserving the actual halo-halo spatial correlations,
the halo mass distribution, and the correlation signal from non-halo
particles in the simulations.  Figure~2 compares the results for the
original output in the $n=-2$ scale-free simulation with the results
for replaced power-law halos.  Figure~3 does the same for the
$\om=0.3$ CDM simulation.  As indicated, when the halo profiles obey
$\rho\propto r^{-\epsilon}$, we find that $\Delta$ and $Q$ at high $k$
are well approximated by power laws, and that the slopes are indeed
related to $\epsilon$ according to equation~(\ref{epsilon}).  For
$\epsilon=2$ (i.e. an isothermal density distribution), for example,
both Figures~2 and 3 show that $\Delta\propto k$ and $Q_{\rm
eq}\propto k$.  Figure~3 also verifies equation~(\ref{epsilon}) for
$\epsilon=2.35$: $\Delta\propto k^{1.7}$ and $Q_{\rm eq}\propto
k^{0.65}$.  We therefore conclude that for a power-law density
profile, equation~(\ref{epsilon}) works well at predicting the slopes
of the power spectrum and the three-point $Q$ in the nonlinear regime.
This remains true even when the clustering of the halo centers and the
correlation signal from non-halo particles are taken into account.

\section{Discussion}

To gain a better understanding of the behavior of two- and three-point
statistics in the deeply nonlinear regime, we have investigated in
this {\it Letter} how the shapes of dark matter halos affect the power
spectrum $P$ and the three-point amplitude $Q$ at high $k$.  For halos
with power-law density profiles, Figures~2 and 3 demonstrate that the
simple model of equation~(\ref{epsilon}) works very well at predicting
the slopes of $P$ and $Q$ at high $k$.  The figures also show that the
converse is not true: the slope of $P$ or $Q$ {\it alone} does not
determine a unique halo shape.  For example, Figure~2 shows that $P$
at high $k$ in the $n=-2$ scale free model is only slightly changed
when simulation halos are replaced with $\epsilon=2$ power-law halos.
It would therefore be difficult to distinguish these two profiles
using the power spectrum alone.  The three-point $Q$, however, is
drastically different for the two profiles.  This indicates that
although the nonlinear $P$ and $Q$ are both closely related to the
shapes of halos, they probe different regions and properties of the
halos.  A plausible explanation for this difference is that the
two-point function reflects the mean density at a given distance from
a particle, whereas the three-point function reflects the dispersion
in density at the given scale (Peebles~1980, \S~36).

Realistic dark matter halos do not have pure power law profiles.
High-resolution $N$-body simulations have shown that cold dark matter
halos have a roughly (Jing \& Suto 2000) universal density profile
with $\epsilon=3$ at the outer radii and $\epsilon=1$ (Navarro, Frenk
\& White 1997; Huss, Jain, \& Steinmetz 1999) or $\epsilon=1.5$ (Moore
et al. 1999) nearer the halo centers.  A further complication is that
these profiles have a mass-dependent ``concentration parameter'': less
massive halos are more centrally concentrated.  It will be useful to
develop a model more extensive than equation~(\ref{epsilon}) to
incorporate these factors.  Such a model should help to shed light on
why $Q(k)$ at high $k$ in our figures is approximately a power law,
even though the halo profile in the simulations is not a pure power
law.  Given that the shape of $Q$ is highly sensitive to halo
profiles, the model can also be used to tackle the question: what
criteria must the halos satisfy in order for hierarchical clustering
to occur?

In the best simulation of the $n=-2$ scale-free model available to us
(with 16.7 million particles and a force resolution of $L/5120$), we
find that the three-point amplitude $Q$ is not a constant with
wavenumber $k$ and that the self-similar scaling observed for the
two-point function (Jain \& Bertshinger 1998) does not hold as
rigorously for the three-point function (Fig.~1).  We have checked for
possible numerical artifacts using smaller simulations with different
phases (see \S~2), and the results are similar within the range of
overlap but may still be a manifestation of the finite simulation
volume, to which the $n=-2$ spectrum is particularly susceptible.
Based on lower resolution simulations for scale-free initial
conditions with spectral index $n$, Fry et al.~(1993) suggested $ Q =
3/(3+n) $ as the asymptotic value in the nonlinear regime.
Scoccimarro \& Frieman (1999) in so-called hyperextended perturbation
theory obtain $Q = (4 - 2^n)/(1 + 2^{n+1})$.  These two expressions
would predict $Q=3$ and 2.5, respectively, for $n=-2$ at high $k$.
Figure~1 shows that $Q$ reaches a similar value between 2.5 and 3 at
$k\sim k_{nl}$, but it rises to 4 to 5 at $ k\sim 10\,k_{nl}$.  For
CDM-type models, the effective index decreases with $k$, so the two
expressions above would give $Q$ rising with $k$, but in neither case
is the prediction as strong as what we see.

It is useful to compare our numerical results with theoretical
expectations.  From the assumptions of stable clustering and
self-similarity, it is well known that for the two-point function,
$\Delta(k)\propto k^{(9+3n)/(5+n)}$ at high $k$ for an initial $
P_l\propto k^n $ (Davis \& Peebles 1977).  The nonlinear $\Delta$ in
Figure~1 and earlier work of Efstathiou et al.~(1988), Peacock \&
Dodds (1996), Jain (1997), Ma (1998), and Jain \& Bertschinger (1998)
all support this behavior.  For the three-point function, these same
conditions would result in constant $Q$, i.e., hierarchical
clustering.  The fact that high-resolution simulation results do not
support this implies that one of these assumptions -- stability or
self-similarity -- must be violated, or still better future simulations 
are needed.

%%%%%%%%%%%%%%%%%%%%%%%%%%%%%%%%%%%%%%%%%%%%%%%%%%%%%%%%%%%%%%%%%%%%%%%%%
\acknowledgments

We are grateful to John Peacock for enlightening discussions and
Edmund Bertschinger for providing the $n=-2$ scale-free simulation.
Computing time for this work is provided by the National Scalable
Cluster Project and the Intel Eniac2000 Project at the University of
Pennsylvania.  C.-P. M. acknowledges support of an Alfred P. Sloan
Foundation Fellowship, a Cottrell Scholars Award from the Research
Corporation, a Penn Research Foundation Award, and NSF grant AST
9973461.
%%%%%%%%%%%%%%%%%%%%%%%%%%%%%%%%%%%%%%%%%%%%%%%%%%%%%%%%%%%%%%%%%%%%%%%%%

\newpage
%%% Figure 1
\begin{figure}
%\epsscale{0.75}
\plotone{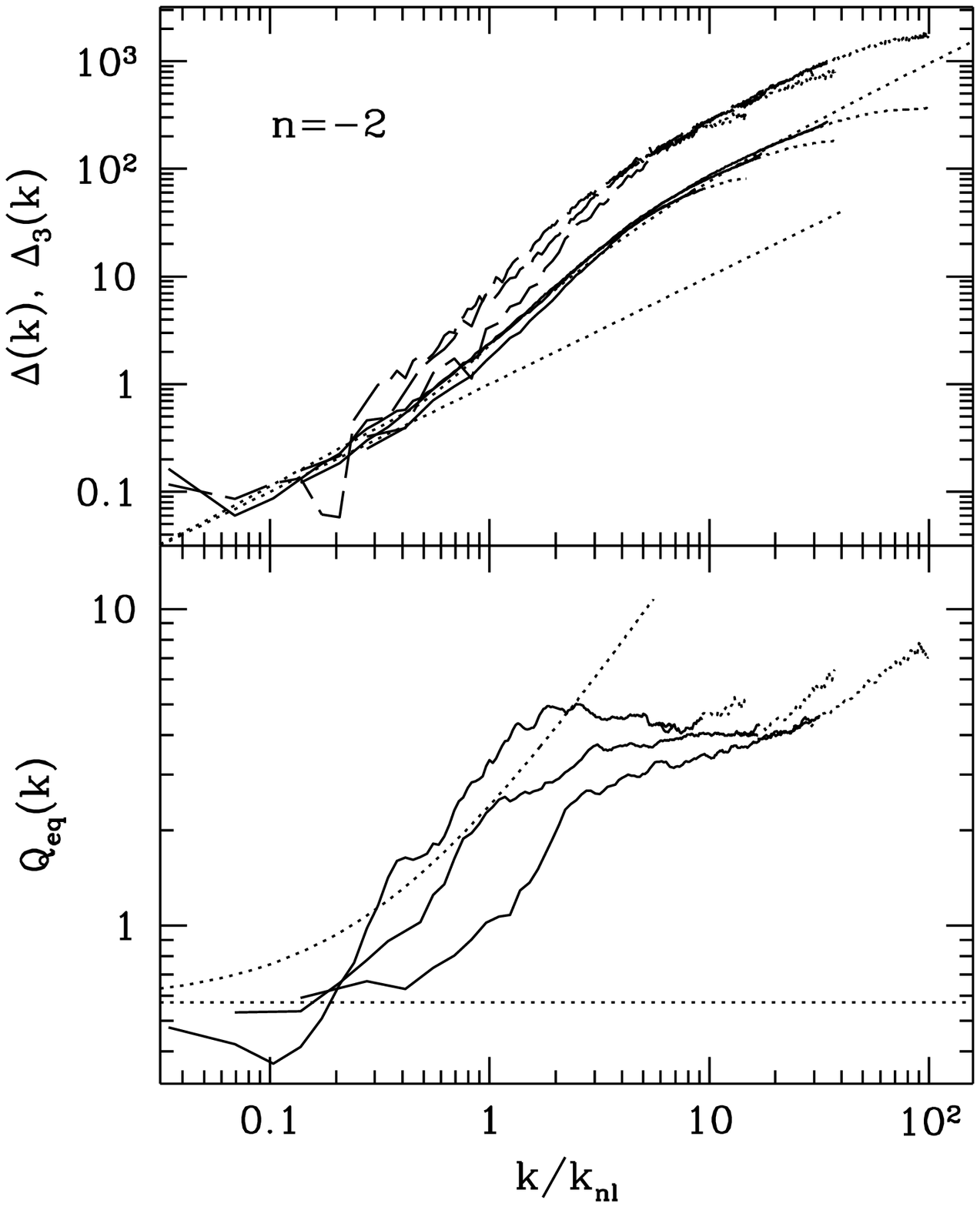}
\caption{Scaling of the power spectrum, $\Delta(k)=4\pi k^3 P(k)$, and
the three-point statistics $\Delta_3(k) = 4\pi k^3\,B^{1/2}_\eq (k)$
and $Q_\eq (k)$ for equilateral triangles, in an $n=-2$ scale-free
simulation.  Three time outputs are shown, where the expansion factor
(1 initially) and nonlinear wavenumber (in units of $2\pi/L$) are:
$(a,k_{nl})= (13.45, 29), (19.03, 14.5)$, and $(26.91, 7.25)$ (left to
right).  In the top panel, the solid and dashed curves are $\Delta$
and $\Delta_3$, respectively, computed from the simulation; the dotted
straight line is the linear $\Delta_l\propto k$; and the dotted curve
is the analytic approximation for the nonlinear $\Delta$ from Peacock
\& Dodds (1996).  In the bottom panel, the dotted curves indicate the
leading (horizontal line) and next-to-leading perturbation theory
results for $Q_{\rm eq}\,$. The dotted portions of the curves at high
$k$ indicate regions of possible spurious effects due to finite
numerical resolution.}
\end{figure}

%%% Figure 2
\begin{figure}
%\epsscale{0.75}
\plotone{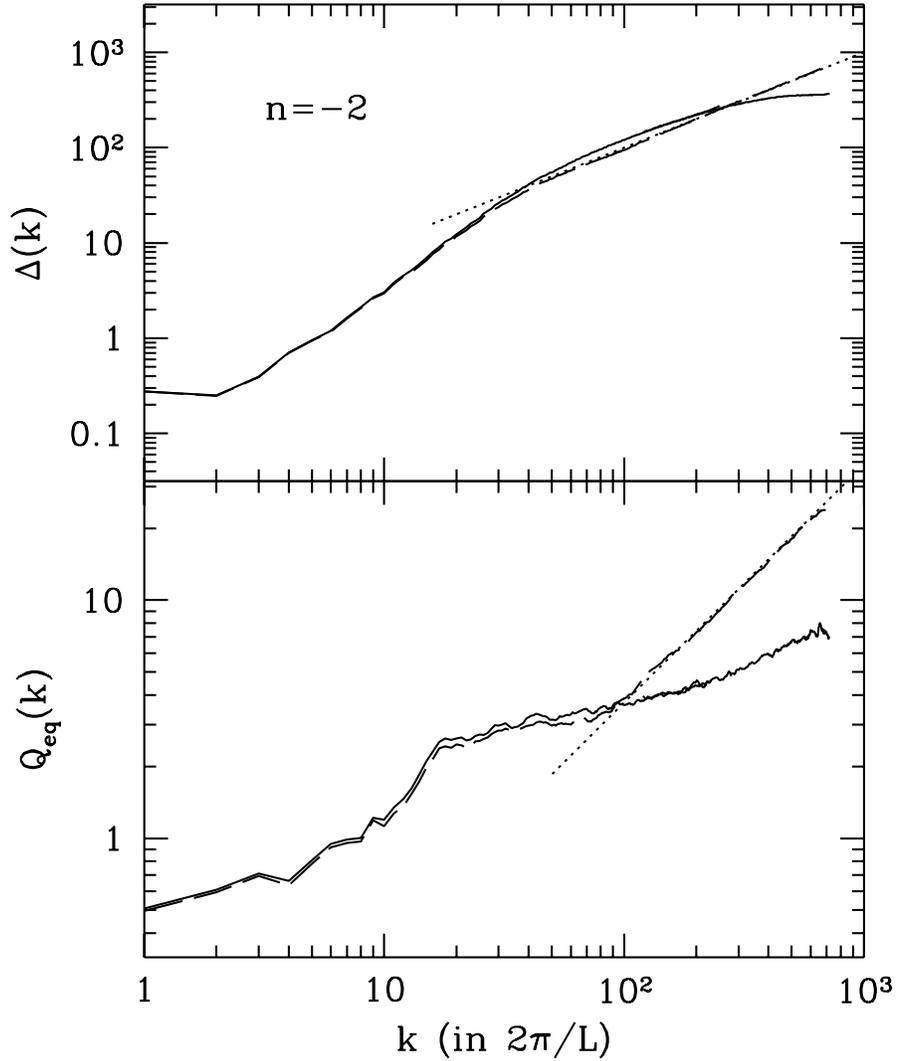}
\caption{The power spectrum $\Delta(k)=4\pi k^3 P(k)$ and the
three-point amplitude $Q_{\rm eq}(k)$ for the last output
($k_{nl}=7.25$) of the $n=-2$ model in Fig.~1.  In both panels, the
solid curves are computed from the original simulation particles.  The
dashed curve compares the results when the simulation particles in
halos are redistributed according to pure power-law $\rho\propto
r^{-\epsilon}$ with $\epsilon=2$.  The straight dotted line marks the
slope predicted by eq.~(\ref{epsilon}), which works remarkably well at
$ k \go 70$.}
\end{figure}

%%% Figure 3
\begin{figure}
%\epsscale{0.75}
\plotone{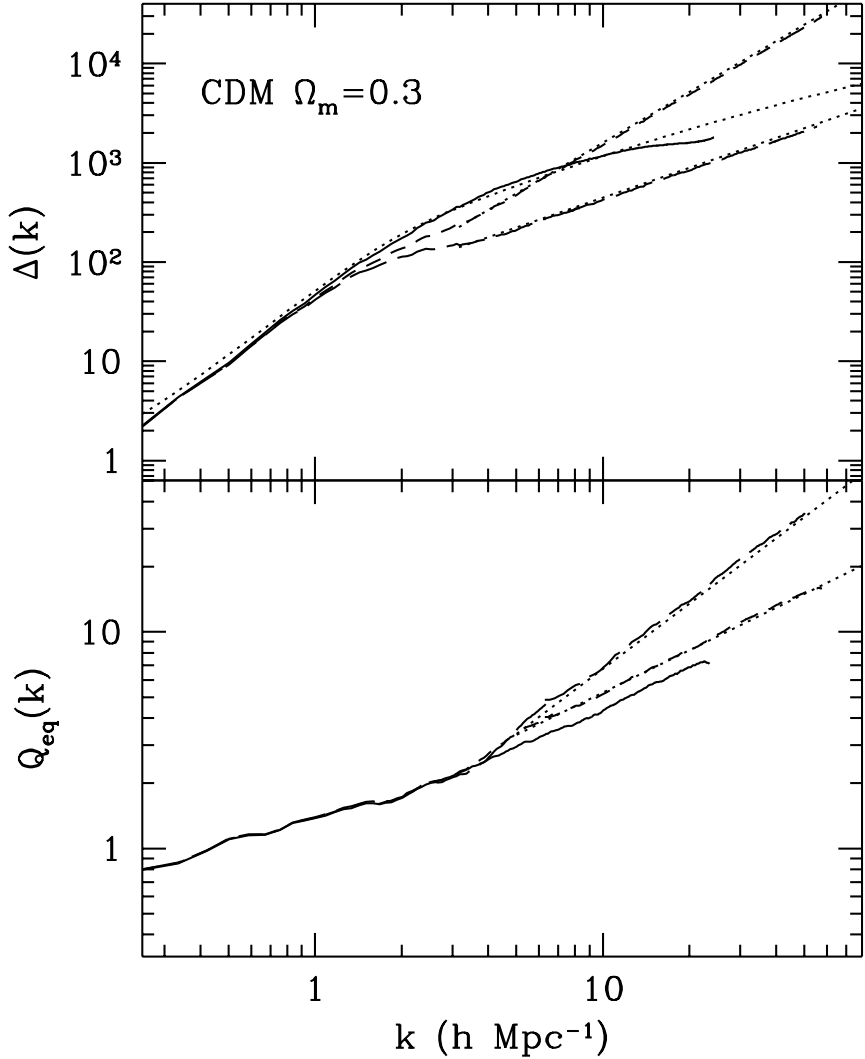}
\caption{Similar to Fig.~2, but for a low-density CDM model with
$\om=0.3$ and $\ov=0.7$.  In both panels, the solid curves are
computed from the original simulation particles.  The dotted curve
tracing the solid one in the top panel shows the analytic
approximation for $\Delta$ from Ma (1998).  The other curves compare
the results when the simulation particles in halos are redistributed
according to pure power-law $\rho\propto r^{-\epsilon}$ with
$\epsilon=2$ (long-dashed) and $\epsilon=2.35$ (short-dashed).  The
straight dotted lines mark the slopes predicted by
eq.~(\ref{epsilon}), which works remarkably well at $ k \go 3 \, h \,
{\rm Mpc}^{-1} $.}
\end{figure}

\end{document}